\documentclass{arXiv}

\title{$^{30}$S RI Beam Production and X-ray Bursts}

\ShortTitle{$^{30}$S RI Beam Production and X-ray Bursts}

\author{\speaker{David Kahl},$^{a}$ Alan A. Chen,$^{a}$ Dam Nguyen Binh,$^{b}$ Jun Chen,$^{a}$ Takashi Hashimoto,$^{b}$ Seiya Hayakawa,$^{b}$ 
Aram Kim,$^{c}$ Shigeru Kubono,$^{b}$ Yuzo Kurihara,$^{b}$ Nam Hee Lee,$^{c}$ Shin'ichiro Michimasa,$^{b}$ Shunji Nishimura,$^{d}$ 
Christian Van Ouellet,$^{a}$ Kiana Setoodeh nia,$^{a}$ Yasuo Wakabayashi,$^{b}$ and Hidetoshi Yamaguchi.$^{b}$\\
\llap{$^{a}$}Department of Physics \& Astronomy, McMaster University, Canada\\
\llap{$^{b}$}Center for Nuclear Study, Graduate School of Science, the University of Tokyo, Japan\\
\llap{$^{c}$}Department of Physics, Ewha Womans University, Korea\\
\llap{$^{d}$}RIKEN (The Institute of Physical and Chemical Research), Japan\\
E-mail: \email{daidxor@gmail.com}}

\abstract{The present work reports the results of $^{30}$S radioactive beam development for a future experiment directly measuring data to extrapolate the $^{30}$S($\alpha$,p) stellar reaction rate in Type I X-ray bursts, a phenomena where nuclear explosions occur repeatedly on the surface of accreting neutron stars.  We produce the radioactive ion $^{30}$S via the $^{3}$He($^{28}$Si,$^{30}$S)n reaction, by bombarding a cryogenically cooled target of $^{3}$He at 400 Torr and 80 K with $^{28}$Si beams of 6.9 and 7.54 MeV/u.  In order to perform a successful future experiment which allows us to calculate the stellar $^{30}$S($\alpha$, p) reaction rate, Hauser-Feshbach calculations indicate we require a $^{30}$S beam of $\sim 10^{5}$ particles per second at $\sim 32$ MeV.  Based on our recent beam development experiments in 2006 and 2008, it is believed that such a beam may be fabricated in 2009 according to the results presented.  We plan to measure the $^{4}$He($^{30}$S,p) cross-section at astrophysical energies in 2009, and some brief remarks on the planned ($\alpha$,p) technique are also elucidated.}
\FullConference{Submitted for the Proceedings of the\\
		 10th Symposium on Nuclei in the Cosmos\\
		 July 27 - August 1 2008\\
		 Mackinac Island, Michigan, USA}
\begin{document}
\section{Motivation}
The $^{30}$S($\alpha$,p)$^{33}$Cl reaction is a significant link in the {\it $\alpha$p} process, which competes with the {\it rp} process on accreting neutron stars \cite{iliadis99, schatz99, anuj}, but there is no experimental data to date.  \underline{X}-\underline{r}ay \underline{b}ursts (XRBs) with multiple peaks in their bolometric luminosity are explained by Fisker, Thielemann \& Wiescher as multiple releases of nuclear 
energy, separated by a nuclear waiting point \cite{fisker04}.  Fisker, Thielemann \& Wiescher's 2004 model reproduces these double-peaks for lower accretion rates ($\sim8 \times 10^{-10} M_{\odot}$ year$^{-1}$) of solar material \cite{anders89} in \underline{l}ow \underline{m}ass \underline{X}-ray \underline{b}inaries (LMXBs), resulting in helium shell flashes.  In all XRBs, pre-burst burning takes place via the $\beta$-limited \underline{h}ot CNO cycle at $T_{9}=0.1-0.2$ K ($T_{9}=10^{9}$ K).  A thin-shell instability leads to explosive nucleosynthesis \cite{vanhorn, hansen75}, leading to break-out from the HCNO cycles, and the {\it rp} process commences primarily on $^{19}$Ne and $^{21}$Na seed nuclei \cite{wiescher99} at $T_{9}=0.4-1.3$ K \cite{woosley04, fisker08}.  The {\it $\alpha$p} process, which does not rely on slow $\beta$-decays, begins to compete with the {\it rp} process, increasing the energy generation rate by as much as three orders of magnitude in one second \cite{wallace81}.

XRBs exhibit recurrent bursts with frequencies typically ranging from hours to days, and there are some 100 known X-ray bursters.  However, three systems have exhibited multiple peaks in their bolometric luminosities \cite{sztajno, paradijs, penninx, kuulkers}, which constrain the models of XRBs, although there have been a plethora of proposed explanations since the were first observed.  Fisker, Thielemann \& Wiescher recently proposed a nuclear waiting point at $^{30}$S, because the modeled burning is impeded from further (p,$\gamma$) captures by ($\gamma$,p) reactions, and by varying the Hauser Feshbach statistical reaction rate of $^{30}$S($\alpha$,p) by a factor of 100, Fisker {\it et al.} produce a different burst profile where the double peak structure is notably diminished \cite{fisker04}.  We will perform a direct experimental measurement of the $^{4}$He($^{30}$S,p) cross section, which requires a low energy ($E_{beam}=32$ MeV) $^{30}$S RI beam of intensity $\sim 10^{5}$ pps and high purity.  We plan to scan the energy region corresponding to the Gamow window from 1--2 GK (1.4 MeV $\leq E_{cm} \leq$ 3.8 MeV).

$\alpha$-capture reactions in the $\alpha$p process proceed through T$_{z}= \frac{(N-Z)}{2}=-1$ compound nuclei, and cross sections on nuclei with T$_{z}= \pm1$ near A$\sim$20 are shown to be dominated by natural parity $\alpha$-cluster resonances \cite{bradfield99a, bradfield99b, groombridge, dababneh, apra05}.  Similar behavior is expected at higher mass regions \cite{angulo99},  casting doubt on the results of Hauser Feshbach statistical modeling in such cases \cite{fisker04}.  With the exception of $^{4}$He($^{18}$Ne,p)$^{21}$Na \cite{bradfield99a, bradfield99b, groombridge}, there is no direct ($\alpha$,p) experimental data involving radioactive nuclei.  These experimental data on $^{18}$Ne($\alpha$,p) are taken mostly at energies higher than those found in XRBs, resulting in a small overlap with the Gamow window.  Charged particle reactions are inhibited by the Coulomb barrier, and the cross sections of typical astrophysical reactions range from below picobarns to the microbarn range \cite{wiescher99}.  The unlikely reaction probabilities require extremely low levels of detector background coupled with long stretches of nuclear beam time in order to get good reaction statistics.  With respect to radioactive nuclei interacting with alpha particles, the foil windows often used for gaseous helium targets pollute the spectra with scattering events and cause energy straggling in the beam and ejecta, and radioactive beam time is a highly sought after commodity which limits the time available for individual experiments.  To optimize use of RI beam time, indirect methods are critical for locating possible resonances, but there is at present only one known level (no spin/parity assignment) in $^{34}$Ar at the relevant astrophysical energies (7.54 MeV $\leq E_{x} \leq$ 9.55 MeV).

\section{$^{30}$S Beam Production}
\begin{figure}
\centering
\includegraphics[angle=0, scale=.4, clip=true, trim=0 280 0 280]{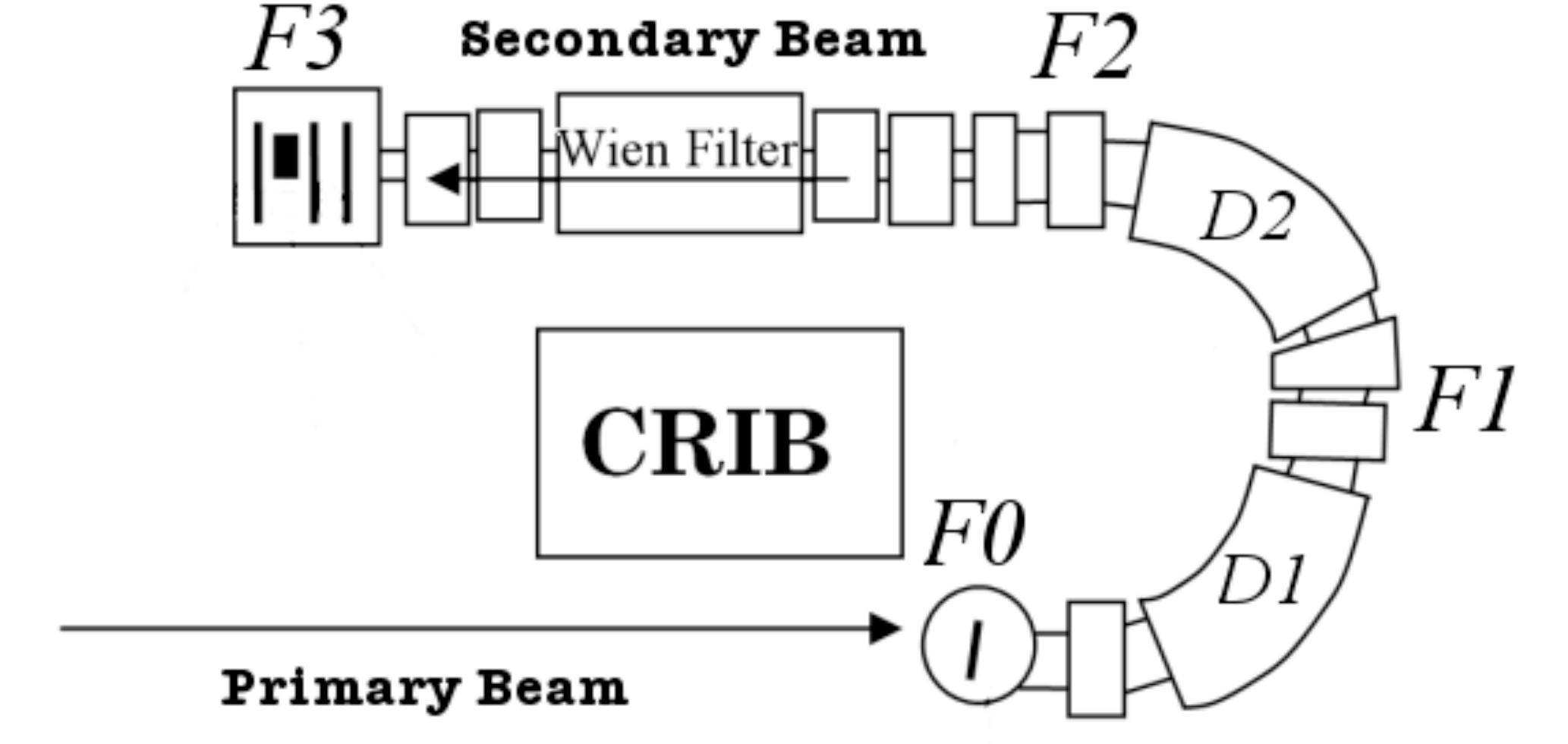} 
\caption[Schematic view of the CRIB separator facility]
{Schematic overhead view of the CRIB separator facility.  The primary beam enters from the lower left side and impinges a cryogenic, windowed gas target at F0.  F1 is the momentum dispersive focal plane.  F2 is a doubly achromatic focal plane.  F3 is the experimental scattering chamber.  D1 \& D2 are magnetic dipoles.  There are 7 magnetic quadrupoles and two magnetic multipoles to control beam optics.  The Wien (velocity) Filter is used to increase RI beam purity.}
\label{crib_fac}
\end{figure}
We successfully separated out the proton-rich isotope $^{30}$S, after bombarding a cryogenic, windowed $^{3}$He gas cell with a stable $^{28}$Si beam using the $^{3}$He($^{28}$Si,$^{30}$S)n reaction.  We tested two primary beams: $^{28}$Si$^{9+}$ (6.9 MeV/u and 100 pnA); $^{28}$Si$^{10+}$ (7.54 MeV/u and 10 pnA); both beams were extracted from the CNS HyperECR ion source and accelerated by the AVF cyclotron of the RIKEN Nishina Center.  We tested three production target gas pressures up to 400 Torr, finding $^{30}$S yield was maximized at the highest pressure we tested.  We conducted two day RI beam development runs in both 2006 and 2008 at the CRIB facility \cite{KUBONO1, YANAGISAWA} (see Figure \ref{crib_fac}), operated by the University of Tokyo and located at RIKEN.  The cocktail beam is sent through two magnetic dipoles to an achromatic focal point F2, where we identify different species in the cocktail beam by radiofrequency ($RF$) time-of-flight ($TOF$) and energy $E$.  We also tested a 1.5 $\mu$m mylar degrader at the dispersive focal plane.

To measure the $^{4}$He($^{30}$S,p)$^{33}$Cl cross section, we will bombard a thick $^{4}$He gas cell with a mono-energetic beam of $^{30}$S.  We will gate on $^{30}$S beam ions in $RF$ and $TOF$ and detect reaction protons downstream.  We will scan the Gamow window for $T_{9}=1-2$ K $\rightarrow$ $E_{beam}=11.3-32.3$ MeV on target.  We then plan to reconstruct event-by-event $^{4}$He($^{30}$S,p) reactions, by tracking $^{30}$S beam ions in two delay-line PPACs, measuring corresponding recoil protons with an array of $\Delta E-E$ silicon telescopes at various angles.  We tested three different $^{4}$He gas cells and detector setups with beams of $^{14}$O, $^{21}$Na, and $^{28}$Si.  We plan to measure the $^{4}$He($^{30}$S,p) cross section in 2009.  

\section{Results}
\begin{figure}
\centering
\includegraphics[angle=0, scale=.33, clip=true, trim=0 30 0 140]{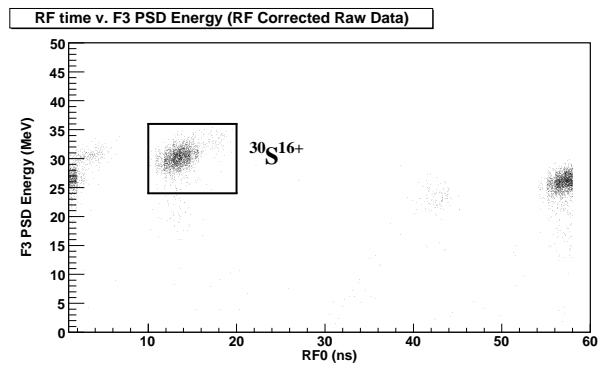} 
\caption[RF vs. SSD Energy raw spectrum at F3 for $^{30}$S$^{16+}$]
{The plot shows $RF~time$ in nanoseconds on the abscissa and PSD energy in MeV on the ordinate for the beam after traversing the CRIB, including both F3 PPACs and a 2.5 $\mu$m Havar target, and the Wien Filter is set to 60 kV in May 2008.  The species of interest, $^{30}$S$^{16+}$, is indicated.  The summed data collection period is 2265 seconds, a total of 8,169 particles are detected, 2458 of which are identified as $^{30}$S$^{16+}$.}
\label{30s16_f3_data}
\end{figure}

\begin{figure}
\centering
\includegraphics[angle=0, scale=.33, clip=true, trim=0 30 0 140]{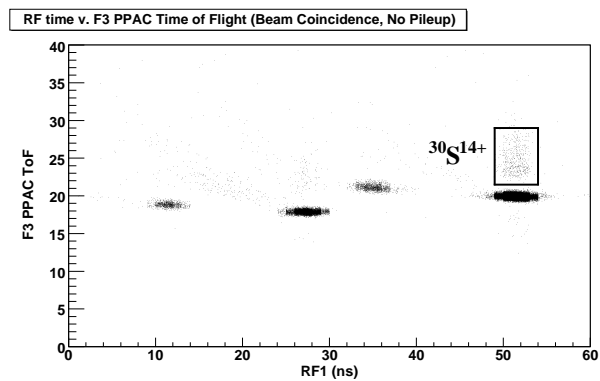} 
\caption[RF vs. ToF raw spectrum at F3 for $^{30}$S$^{14+}$]
{The plot shows $RF~time$ in nanoseconds on the abscissa and F3 PPAC $ToF$ in nanoseconds on the ordinate for the beam after traversing the CRIB, and the Wien Filter is set to $\sim$60 kV in May 2008.  The species of interest, $^{30}$S$^{14+}$, is indicated.  The data collection period is 640 seconds, a total of 93,714 particles are detected, 586 of which are identified as $^{30}$S$^{14+}$.}
\label{30s14_f3_data}
\end{figure}

All our $^{30}$S RI beam results were improved with use of $^{28}$Si$^{10+}$ (7.54 MeV/u) primary beam.  $^{30}$S$^{16+}$ achieved 30\% purity, 30 MeV, and 500 pps per 10 pnA at F3.  $^{30}$S$^{14+}$ achieved 0.8\% purity, 32 MeV, and 1.2 $\times 10^{4}$ pps per 10 pnA at F3.    We determined that the $RF$-$TOF$ spectrum of $^{30}$S$^{15+}$ is contaminated by the leaky beam $^{28}$Si$^{14+}$ due to similar charge-to-mass ratios, although we did not attempt to separate these two ions with a thin degrader at the momentum dispersive focal plane F1.  Figure \ref{30s16_f3_data} shows $^{30}$S$^{16+}$ is easy to distinguish, but the intensity is far below the desired 10$^{5}$ pps, however, the WF transmission was an astonishing 80\%.  Figure \ref{30s14_f3_data} shows we separate $^{30}$S$^{14+}$ in the $RF$ vs. $TOF$ spectrum at $\sim$ 10$^{4}$ pps per 10 pnA, but due to limited time, we were not able to optimize the purity.  When we attempted the use of a 1.5 $\mu$m aluminized mylar degrader at F1, the intensity on-target decreased by 80\%, which can be improved with better optics tuning.

In 2008, we also tested the charge-state distribution of $^{28}$Si beam ions in carbon compared to Havar foil, to test if adding a carbon stripper foil after our Havar-windowed production target increases the yield of highly charged states of $^{30}$S.  Although we attempted to measure the charge-state distributions directly with $^{30}$S in May 2008, the primary beam current was fluctuating greatly due to instability of the HyperECR, and we were unable to properly normalize the data for comparison.  However, for a $^{28}$Si beam of 3.4 MeV/u (comparable to the $^{30}$S energy achieving maximum transmission intensity at this point in the beam-line), it was found that $^{28}$Si$^{14+}$ transmission through carbon foil is an order of magnitude higher than for Havar foil, $^{28}$Si$^{13+}$ transmission was a factor of 5 greater for carbon compared to Havar, and $^{28}$Si$^{12+}$ transmission was less than a factor of 2 greater for carbon compared to Havar.  Thus, it may be possible to preferentially populate higher charge states of $^{30}$S by use of a carbon stripper foil.

In the future, we hope to achieve the requisite 10$^{5}$ pps by extracting $^{28}$Si$^{10+}$ at a higher rate allowing for intensities of up to 30 pnA at F0, and accelerate the beam for a larger number of turns in the AVF cyclotron up to 8.2 MeV/u.  By comparing the $^{30}$S intensity results from the $\sim 0.5$ MeV/u energy increase from Decemeber 2006 to May 2008, it is concluded that a 0.5 MeV/u energy increase in this range corresponds to nearly an order of magnitude increase in the $^{30}$S production yield.  The production gas target has successfully held up to 700 Torr, so we may test if RI beam yield continues to increase with pressure.  By combining these various improvements, we expect to achieve a $^{30}$S intensity above 10$^{5}$ pps in the future.

This work is made possible through the CNS and RIKEN collaboration and funding from the National Science and Engineering Research Council of Canada.

\end{document}